\documentclass[final,aps,prb,twocolumn,superscriptaddress,showpacs]{revtex4-1}

\usepackage{graphicx}
\usepackage{dcolumn}
\usepackage{amsmath,bm}
\usepackage{color}%
\usepackage{multirow}
\usepackage{array} 
\usepackage{esvect}
\usepackage{braket}
\usepackage{makecell}
\usepackage{mhchem}
\usepackage[utf8]{inputenc}
\usepackage{kotex}
\usepackage{xr}
\usepackage{soul,xcolor}
\setstcolor{red} 

\usepackage{xr}

\makeatletter
\newcommand*{\addFileDependency}[1]{
  \typeout{(#1)}
  \@addtofilelist{#1}
  \IfFileExists{#1}{}{\typeout{No file #1.}}
}
\makeatother

\newcommand*{\myexternaldocument}[1]{%
    \externaldocument{#1}%
    \addFileDependency{#1.tex}%
    \addFileDependency{#1.aux}%
}

\myexternaldocument{SI}  

\begin{document}

\title{A new critical growth parameter and mechanistic model for SiC nanowire synthesis via Si substrate carbonization: the role of H$_2$/CH$_4$ gas flow ratio}

\author{Junghyun Koo}
\affiliation{Department of Physics and Research Institute for Basic Sciences, Kyung Hee University, Seoul, 02447, Republic of Korea}

\author{Chinkyo Kim}
\email[Corresponding author. E-mail: ]{ckim@khu.ac.kr}
\affiliation{Department of Physics and Research Institute for Basic Sciences, Kyung Hee University, Seoul, 02447, Republic of Korea}
\affiliation{Department of Information Display, Kyung Hee University, Seoul, 02447, Republic of Korea}


\begin{abstract}
SiC structures, including nanowires and films, can be effectively grown on Si substrates through carbonization. However, growth parameters other than temperature, which influence the preferential formation of SiC nanowires or films, have not yet been identified. In this work, we investigate SiC synthesis via Si carbonization using methane (CH$_4$) by varying the growth temperature and the hydrogen to methane gas flow ratio (H$_2$/CH$_4$).  We demonstrate that adjusting these parameters allows for the preferential growth of SiC nanowires or films. Specifically, SiC nanowires are preferentially grown when the H$_2$/CH$_4$ ratio exceeds a specific threshold, which varies with the growth temperature, ranging between 1200$^\circ$C and 1310$^\circ$C. Establishing this precise growth window for SiC nanowires in terms of the H$_2$/CH$_4$ ratio and growth temperature provides new insights into the parameter-driven morphology of SiC. Furthermore, we propose a mechanistic model to explain the preferential growth of either SiC nanowires or films, based on the kinetics of gas-phase reactions and surface processes. These findings not only advance our understanding of SiC growth mechanisms but also pave the way for optimized fabrication strategies for SiC-based nanostructures.
\end{abstract}
\maketitle

\section{Introduction}
SiC is a semiconductor material distinguished by its wide band gap, high thermal conductivity, and exceptional mechanical strength, making it suitable for high-power and high-temperature applications. Over 200 SiC polytypes exist, including hexagonal ($\alpha$-SiC) and cubic ($\beta$-SiC or 3C-SiC) structures.\cite{Zekentes-JPDAP-44-133001} Among these, 3C-SiC is particularly favored due to its lower lattice mismatch with silicon, which facilitates simpler fabrication processes, reduces costs, and enhances electron mobility. These features make 3C-SiC ideally suited for high-speed electronic and optoelectronic applications.\cite{Zekentes-JPDAP-44-133001,Fraga-MicroM-11-799}

SiC thin films are essential for high-performance electronic devices, especially in power electronics and microelectromechanical systems (MEMS). Additionally, due to their hardness and chemical stability, SiC films are employed as protective coatings, significantly extending the lifespan of components under extreme conditions.\cite{Gupta-BMS-27-445,Zekentes-JPDAP-44-133001,Fraga-MicroM-11-799} The significance of SiC is equally notable when formed as nanowires, offering unique properties for specific applications. The exceptional electronic, optical, and mechanical properties of SiC nanowires, arising from their high surface-to-volume ratio and one-dimensional structure, make them highly suitable for nanoscale devices, sensors, and photodetectors.\cite{Liu-RSCAdv-6-24267,Fraga-MicroM-11-799}

Growth conditions for both SiC films and nanowires have been extensively studied, with various synthesis methods available. Each method requires a systematic variation of parameters such as temperature, precursor flow rates, and substrate types to establish specific growth regimes for SiC nanowires and films. Identifying these growth conditions is crucial not only for achieving desired properties like crystal structure, surface morphology, and defect densities\cite{Gupta-BMS-27-445,Liu-RSCAdv-6-24267,Zekentes-JPDAP-44-133001} but also for enhancing our understanding of the distinct growth mechanisms associated with each method.

Unlike traditional methods that rely on metal catalysts or separate sources for silicon and carbon, catalyst-free carbonization of a Si substrate at high temperatures using methane as the sole carbon precursor provides a straightforward method for synthesizing SiC structures such as nanowires and films. This approach eliminates complications associated with catalyst-induced growth, such as contamination or unwanted reactions. However, growth parameters other than temperature for Si carbonization have not been fully explored to identify conditions that favor the preferential growth of either SiC nanowires or films. In this work, we systematically investigated the role of the H$_2$/CH$_4$ gas flow ratio in this carbonization method and developed a mechanistic model to explain the gas-flow-ratio-dependent preferential formation of SiC, whether as nanowires or films.

\section{Experimental}

SiC nanowires and films were synthesized by carbonizing Si substrates in a ceramic reactor within a furnace. To prevent substrate degradation, all experiments were conducted at growth temperatures ranging from 1200$^\circ$C to 1325$^\circ$C, well below the nominal melting point of Si at 1410$^\circ$C. The furnace temperature was increased at a rate of 23.3$^\circ$C per minute. This was followed by an annealing process lasting 10 to 20 minutes in an atmosphere of argon (Ar) and hydrogen (H$_2$) to remove the native oxide layer and promote optimal surface conditions for SiC nucleation. Subsequently, the carbonization of Si was performed over 10 minutes with variations in the H$_2$/CH$_4$ ratio to systematically investigate its effect on the growth regime of SiC. After the growth phase, the reactor was purged and cooled under an Ar atmosphere to stabilize the SiC structures and prevent further reactions during the cooling process.

\section{Results and discussion}

\subsection{Detailed characterization of synthesized SiC nanowires}

\begin{figure}
\includegraphics[width=1.0\columnwidth]{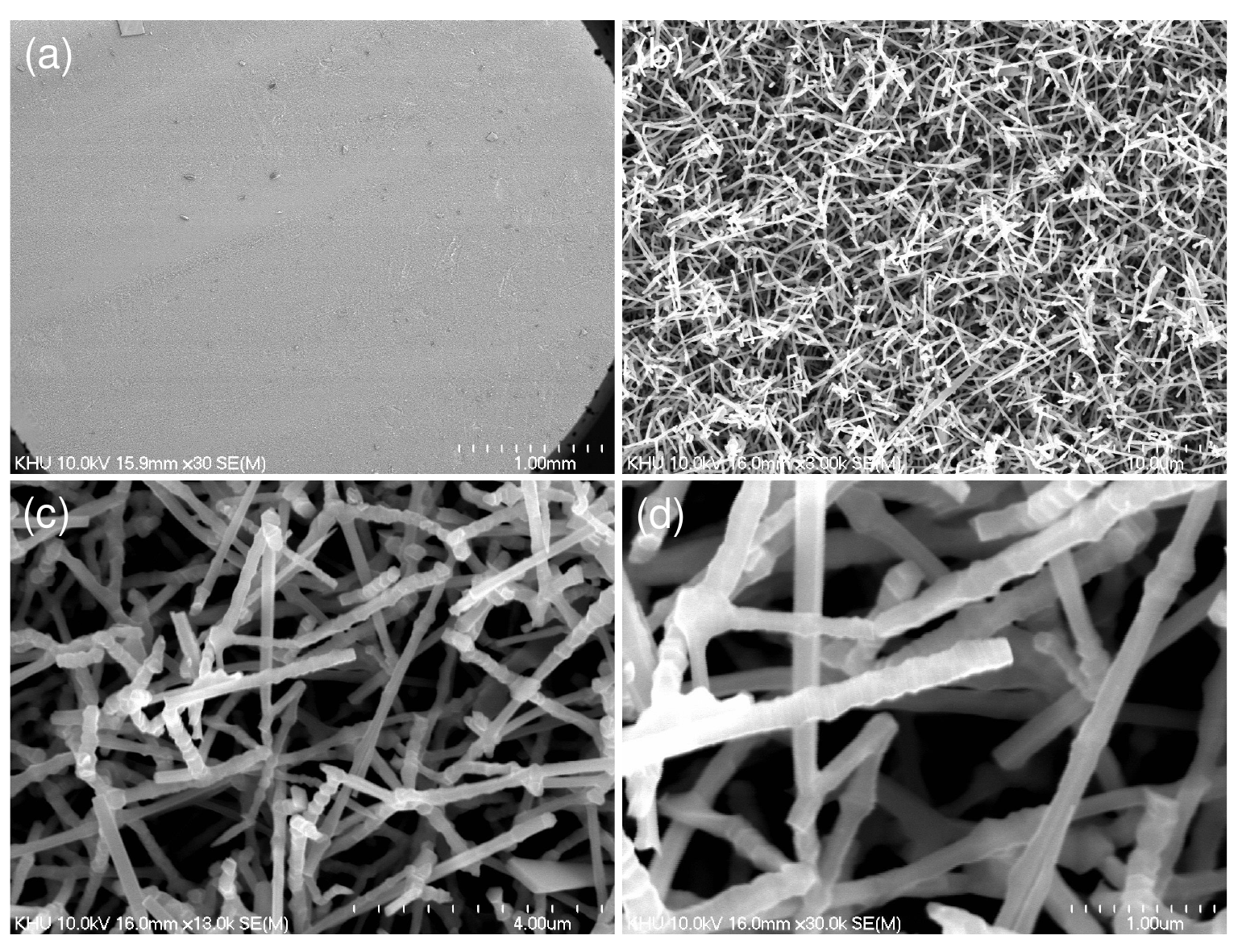}
\caption{SEM images of nanowires grown on Si substrates with CH$_4$ flown at 1300$^\circ$C in different magnifications. Nanowires were uniformly grown across the Si substrate.}
\label{SEM}
\end{figure}

Previous studies have established that SiC nanowires typically grow within a temperature range of 1200$^\circ$C to 1400$^\circ$C, with 1300$^\circ$C identified as the most optimal temperature for growth.\cite{Zekentes-JPDAP-44-133001, Liu-RSCAdv-6-24267, Fraga-MicroM-11-799, Prakash-CN-8-161} Liu \textit{et al.} specifically underscore that a growth temperature of 1300$^\circ$C is ideal for SiC nanowire synthesis.\cite{Liu-RSCAdv-6-24267}

We first present experimental data confirming that the structures grown on Si substrates at a growth temperature of 1300$^\circ$C and an H$_2$/CH$_4$ flow ratio of approximately 100 are indeed SiC nanowires. These nanowires, ranging in length from several tens to hundreds of microns and in diameter from 50 to 100 nm, were uniformly grown across the substrate. Fig.~\ref{SEM} shows a series of progressively zoomed-in SEM images of the nanowires obtained under this condition, showcasing their thick and uniform growth.  The absence of metallic droplets at the nanowire tips indicates a vapor-solid (VS) growth mechanism, rather than the vapor-liquid-solid (VLS) mechanism noted in previous studies.\cite{Huang-Cgd-13-10,Huang-RSCA-4-18360}  

\begin{figure}
\includegraphics[width=1.0\columnwidth]{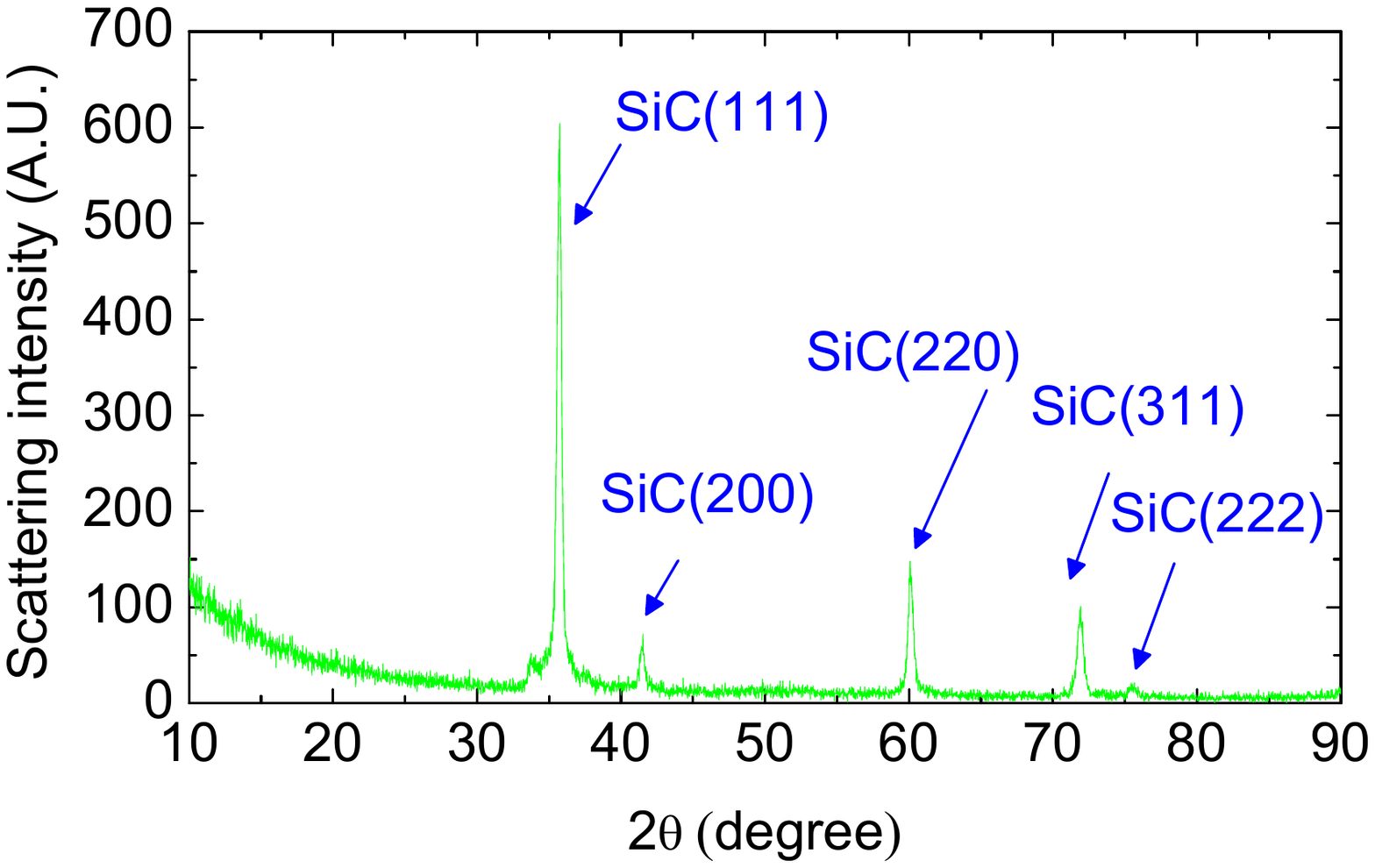}
\caption{The intensity of grazing incidence X-ray diffraction of one of the nanowire samples.  Bragg peaks are well matched to those peaks of 3C-SiC.  Note that no substrate peak is shown due to the very small incidence angle of the incoming X-ray.}
\label{GIXRD}
\end{figure}

These nanowires were confirmed 3C-SiC through grazing incidence X-ray diffraction (GIXRD) analysis. Fig.~\ref{GIXRD} displays the GIXRD intensity pattern of a nanowire sample, where the clearly identified Bragg peaks confirm the 3C-SiC composition. Various Bragg peaks in the GIXRD pattern can indicate either vertically oriented nanowires without a preferred crystallographic orientation along the axial direction or randomly oriented nanowires with a specific axial crystallographic orientation. Our SEM images, which show nanowires grown in random directions relative to the Si substrate, suggest that these nanowires have a distinct crystallographic orientation along their axial direction. It is important to note that the Si substrate peak is absent in the GIXRD pattern because the scan was not conducted along the surface normal direction.

\begin{figure}
\includegraphics[width=1.0\columnwidth]{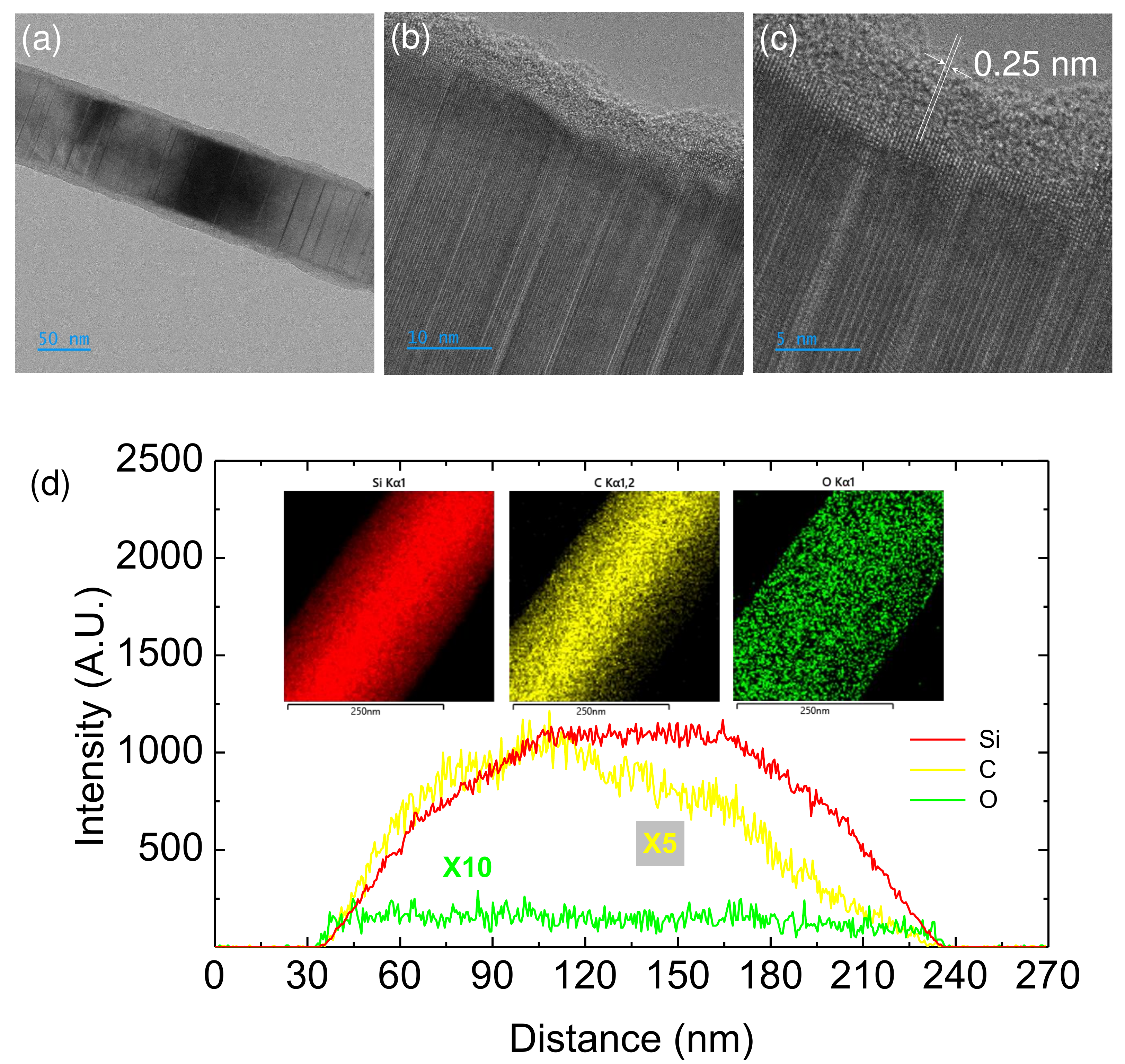}
\caption{(a)$\sim$(c) HR-TEM images of SiC nanowires.  (d)EDS maps and line profiles of SiC nanowires. The constant intensity of O across the nanowire in the line profile suggests that the amorphous shell is SiO$_x$.}
\label{TEM_EDS}
\end{figure}

The preferred orientation of SiC along the axial direction was identified by high-resolution transmission electron microscopy (HR-TEM) image.  Fig.~\ref{TEM_EDS}(a)$\sim$(c) show HR-TEM images of SiC nanowires.  The spacing between consecutive atomic planes along the axial direction is 0.25~nm, which is equal to that of (111) planes of 3C-SiC.  This result suggests that 3C-SiC nanowire was preferentially grown along [111] orientation.\cite{Huang-RSCA-4-18360, Chen-PhscB-42-2335, Wang-Nanotech-19-215602}  Additionally, twin boundary defects were identified, and the dimensions of the nanowires align with those reported for different growth methods.\cite{Huang-RSCA-4-18360, Chen-PhscB-42-2335}  In fact, the TEM image shows an outer shell in an amorphous phase.  Energy dispersive spectroscopy (EDS) was carried out to identify the chemical composition of the nanowire core as well as the outer shell.  Fig.~\ref{TEM_EDS}(d) shows EDS maps and line profiles along the radial direction of the nanowire sample. Consistent with XRD and TEM, the core region is shown to be made of Si and C.  On the other hand, the constant intensity of oxygen across the nanowire along the radial direction suggests that the outer shell is expected to be SiO$_x$, which is reported to be commonly observed for SiC nanowires.\cite{Chen-PhscB-42-2335, Wei-JCG-335-160, Rizk-Drm-17-1660, He-CEC-21-4740, Dong-MatC-103-37,Zekentes-JPDAP-44-133001}

\subsection{Growth regimes of SiC depending on H$_2$/CH$_4$ ratio and growth temperature: nanowire vs film}

\begin{figure*}
\includegraphics[width=1.9\columnwidth]{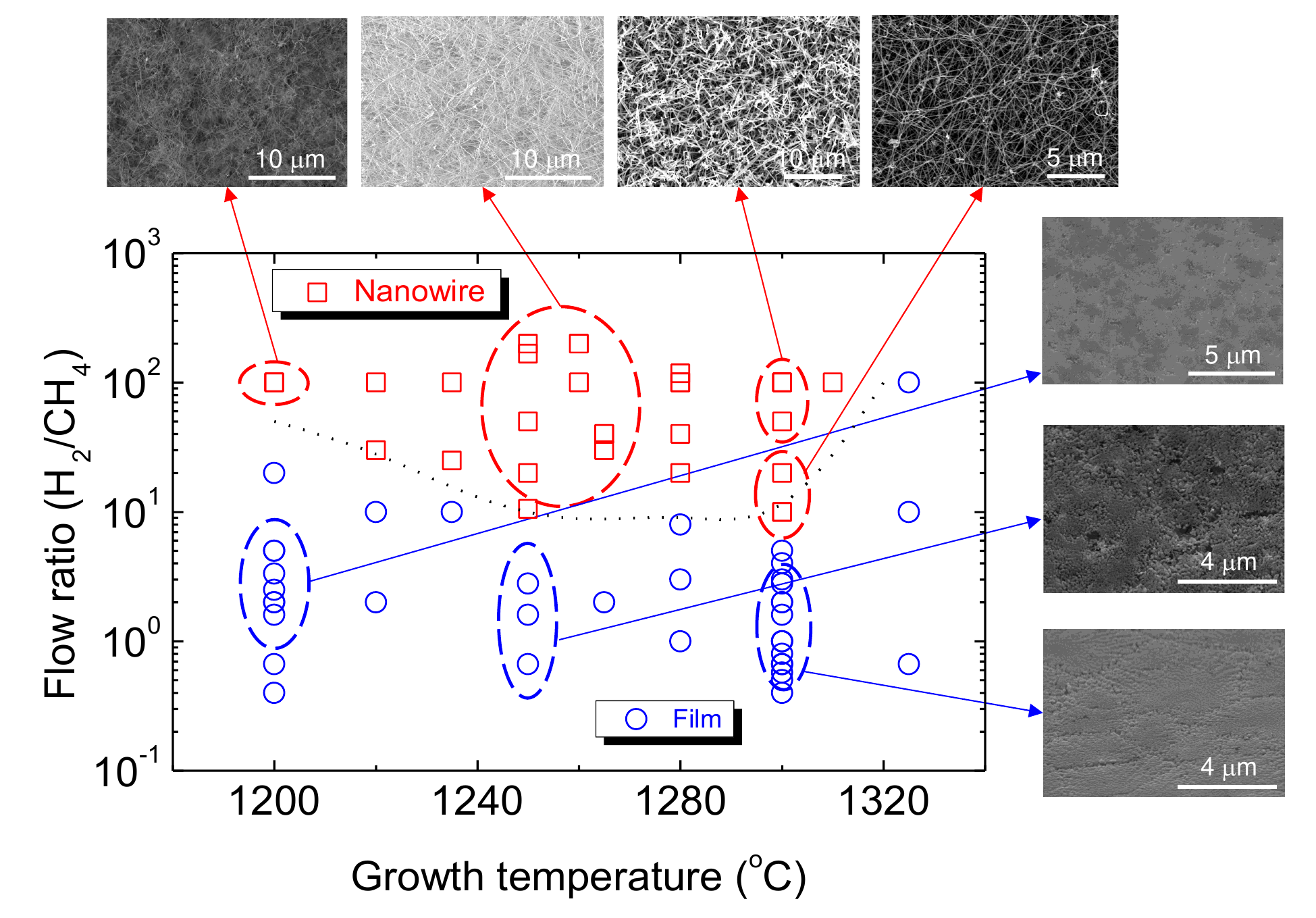}
\caption{Growth regimes of SiC: influence of gas flow ratio and growth temperature on nanowire and film formation on a Si substrate.  We observed both curved and straight configurations of SiC nanowires, with variations primarily influenced by the temperature.  Notably, the curved nanowires resemble those discussed by He and Liu \cite{He-CEC-21-4740, Liu-RSCAdv-6-24267}, while the straight nanowires closely mimic the findings of Wu and Huang.\cite{Wu-Nanotech-19-335602,Huang-RSCA-4-18360} The predominance of either straight or curved nanowires was dependent on the temperature and the H$_2$/CH$_4$ flow ratio.}
\label{Growth-map}
\end{figure*}

Apart from the previous report on the ideal growth temperature (1300$^\circ$C) of SiC nanowires\cite{Liu-RSCAdv-6-24267} and our results described in the previous section, SiC films instead of SiC nanowires were observed to be grown even at the growth temperature of 1300$^\circ$C if H$_2$/CH$_4$ ratio is much smaller than 10.  Furthermore, SiC nanowires were found to be grown at this growth temperature only if the H$_2$/CH$_4$ ratio is larger than a threshold of 10.  At different growth temperatures between 1200$^\circ$C and 1310$^\circ$C, a similar behavior was observed while this threshold changed with the growth temperature.  Extensive data sets on the growth conditions for SiC nanowires and films are plotted in fig.~\ref{Growth-map}.  

This observation underscores the significant interplay between the gas flow ratio and growth temperature, and both influence the formation of SiC structures on Si substrates, as clearly illustrated in Fig.~\ref{Growth-map}. A key insight from this figure is that both the growth temperature and the ratio of H$_2$ to CH$_4$ flow rates are crucial for determining whether SiC nanowires or films are preferentially formed. It suggests that the flow rate of H$_2$ must be at least ten times that of CH$_4$ for nanowires to be preferentially grown, provided that the growth temperature is also within the suitable range for SiC nanowire formation. Thus, both the gas flow ratio and the growth temperature need to be carefully adjusted to successfully grow SiC nanowires on Si substrates.

So far, there have been no prior experimental studies that directly investigate the influence of hydrocarbon partial pressure or the H$_2$/CH$_4$ gas ratio on the growth of SiC nanowires. The lack of this information in this area of research highlights the novelty and importance of our study, which can provide a framework for optimizing the synthesis of SiC nanowires by carbonizing Si substrates with CH$_4$. Furthermore, based on these growth conditions, we can better understand the growth mechanism, as explained in the next section. A detailed discussion on how the formation of either nanowires or films depends on the gas flow ratio will be presented in the following section, along with mechanistic model-based interpretations. 

\subsection{Growth mechanism and the mechanistic model}

Extensive research has focused on the carbonization process involved in our experiment, which is fundamental for forming a buffer layer crucial for the heteroepitaxy of SiC or nitride materials on Si substrates.\cite{Ferro-CRSMS-40-56} It has been suggested that during the initial stage of carbonization, the diffusion of silicon leads to the formation of small voids and nuclei of SiC on the substrate. These nuclei arise from the reaction of carbon from the gaseous precursor with the silicon surface within localized regions.\cite{Severino-JAP-102-2, Li-JES-142-634} As the process progresses, these SiC nuclei begin to expand laterally, adopting an island-like morphology. The islands enlarge and, as the unreacted Si surface area diminishes, some islands start to coalesce, covering the voids and gradually forming a continuous film. This island growth stage is pivotal as the SiC layer transitions from a rough to a more planar, continuous film. The dynamics of SiC nuclei formation on Si substrates are highly dependent on the experimental conditions set during the carbonization process.\cite{Severino-JAP-102-2, Li-JES-142-634}

Regarding the SiC films produced by carbonizing a Si substrate by using CH$_4$ as a carbon precursor, it should be noted that the rough surfaces of these films were commonly observed. Challenges in attaining a smooth SiC surface using CH$_4$ as the carbon precursor have been notable. Only a limited number of studies (approximately two to three) have focused on carbonization with CH$_4$, suggesting that this precursor may have been largely overlooked in early research.\cite{Stinespring-JAP-65-1733, Ferro-JAP-80-4691,Steckl-MRSOPL-242-537} Ferro \textit{et al.} report that CH$_4$ exhibits lower reactivity and cracking efficiency compared to other hydrocarbons such as propane (C$_3$H$_8$). The strong C-H bonds in CH$_4$ result in fewer reactive CH$_x$ radicals, which are necessary for effective SiC film growth. This lower reactivity leads to poorer surface coverage and an overall lower quality of the SiC layer.\cite{Ferro-JAP-80-4691} Additionally, CH$_4$ typically yields a rough surface morphology characterized by the formation of pyramid-like microcrystals, indicative of a three-dimensional (3D) growth mode. This mode is less desirable compared to the smoother, two-dimensional (2D) growth achieved with other precursors, leading to surface heterogeneity and roughness that make CH$_4$ less suitable for producing high-quality SiC films. However, the reactivity profile of CH$_4$ makes it more suitable for the growth of SiC nanowires, where the 3D growth structure is crucial in achieving the desired nanowire morphology and properties.  

To comprehend the flow ratio dependence of SiC nanowires and film growth regime, it is critical to understand: (1) the changes in the nucleation stage when varying the flow ratio, and (2) how these changes influence the growth regime shift between nanowire and film growth. Although prior studies have not explicitly reported on these dynamics across both SiC nanowire and film growth regimes, relevant findings have been observed in the growth of SiC films.  Research by Li and Steckl has shown that increasing the hydrocarbon partial pressure reduces the size of SiC nuclei while simultaneously increasing their density.\cite{Li-JES-142-634}  At low hydrocarbon partial pressures, nuclei tend to form as discontinuous islands, but higher hydrocarbon pressures result in smaller, more densely spread nuclei, which facilitate the formation of a more continuous, film-like structure.\cite{Li-JES-142-634} This leads to the development of SiC films with smoother surfaces and fewer hillocks, indicative of higher quality, under conditions of elevated carbon source partial pressure.\cite{Li-JES-142-634,Steckl-APL-60-2107,Mogab-JAP-45-1075} Similarly, experiments with CO gas as a carbon precursor have demonstrated that the density of SiC nuclei is proportional to the CO partial pressure during the kinetic growth phase. However, as carbonization progresses, the process becomes more thermodynamically driven; lower (higher) CO partial pressures lead to the out-diffusion of Si, resulting in larger (smaller) SiC domains, fewer (greater) nuclei, and a thicker (thinner) SiC film with increased (reduced) surface roughness.\cite{Deura-ASS-660-159965}

As for the second issue of how changes in flow ratios could be responsible for shifting growth regimes, insights can be gleaned from prior work on SiC nanowire synthesis using dual precursors for both carbon and silicon. Studies utilizing methyltrichlorosilane (MTS, CH$_3$SiCl$_3$) as a precursor have revealed interesting behaviors.\cite{Prakash-CN-8-161, Prakash-PSC-43-98, Fu-MCP-100-1, Lien-Cgd-10-36,Song-CR-31-43} When decomposed at high temperatures in the presence of hydrogen, MTS yields CH$_4$ and SiCl$_2$, which subsequently react to form SiC.  These studies have demonstrated that decreasing the H$_2$/MTS ratio increases the merged domain size of nuclei.  In other words, a higher ratio favors the formation of high-quality nanowires and a lower ratio leads to the growth of micro-sized particles or bulky SiC structures.\cite{Prakash-CN-8-161, Prakash-PSC-43-98, Fu-MCP-100-1} This suggests that the size and quality of SiC structures are influenced by specific gas flow ratios, indicating a direct relationship between the size of nuclei and the resulting growth regime. Although these results were derived from experiments not involving the carbonization of Si, they provide valuable insights into how variations in nuclei size might affect the growth regime during SiC synthesis by carbonization.

\begin{figure}
\includegraphics[width=0.9\columnwidth]{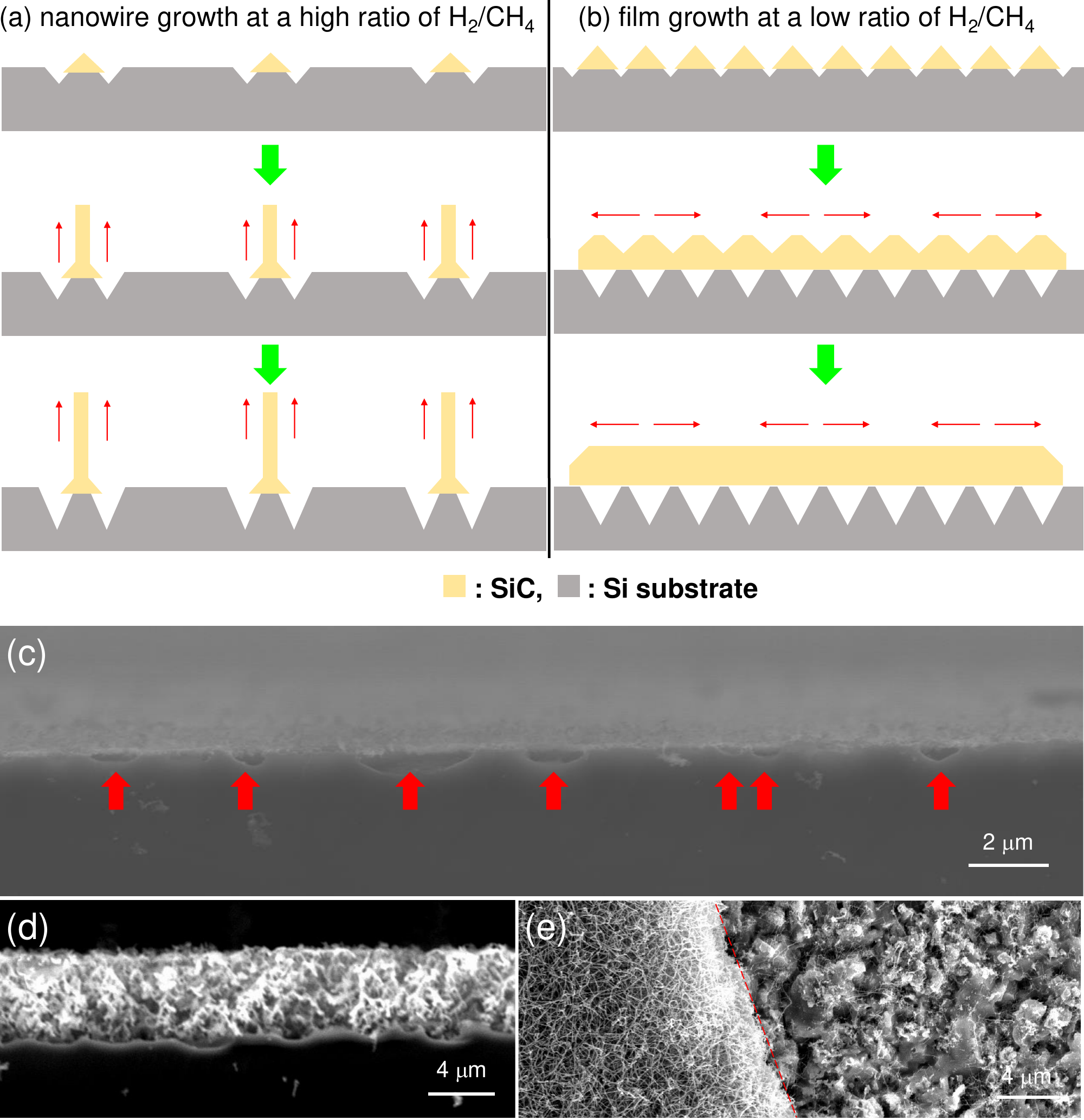}
\caption{Schematic diagrams of growth mechanisms for different growth regimes: (a) nanowires vs (b) films of SiC with H$_2$/CH$_4$ gas flow ratio.  (c) Tilt-view SEM showing pits (marked by red arrows) of a Si substrate.  (d) Cross-sectional SEM image of SiC grown on a Si substrate.  (e) Plan-view SEM image showing SiC nanowires and an exposed Si substrate; the SiC nanowires were self-separated from the substrate on the right half, as indicated by the red dashed line. Note that there is no sign of a SiC film in the exposed Si substrate where SiC nanowires were self-separated from the substrate.}
\label{schematic}
\end{figure}

Thus, combining the arguments given above [(1) the influence of flow ratio change on the nucleation stage of SiC \textit{film} growth and (2) growth regime shift in dual precursor growth situation], this dependence of surface smoothness of SiC films on the partial pressure of hydrocarbon has led us to develop a model to interpret our results on the gas flow ratio dependence of SiC nanowire growth from the kinetic viewpoint.  The schematic diagram of our developed model to interpret the gas flow dependence is illustrated in fig.~\ref{schematic}(a) and (b). In our study, as shown in Fig.~\ref{Growth-map}, it was observed that across most temperature ranges, lower H$_2$/CH$_4$ ratios, corresponding to higher CH$_4$ partial pressures, led to the formation of films.  Conversely, at higher H$_2$/CH$_4$ ratios, which correspond to lower CH$_4$ partial pressures, only nanowires were formed.  These results suggest that under conditions of higher H$_2$/CH$_4$ ratios, the SiC nuclei form as sparse islands, where the vapor-solid (VS) growth mechanism, in which the nuclei act as self-catalysts, dominates over the lateral spreading of the nuclei.\cite{Zekentes-JPDAP-44-133001,Zhang-JPD-42-035108,Wang-MSER-60-1, Huang-RSCA-4-18360} On the other hand, under lower H$_2$/CH$_4$ ratios, the SiC nuclei spread laterally in a dense, film-like pattern, leading to the formation of a continuous SiC film.  This indicates that the process of lateral nucleation and growth is more dominant under these conditions, favoring film formation over nanowire growth.

It is worthwhile to mention a few things on the schematic diagram of our proposed model.  First, there are pits or holes induced on the Si substrate surface.  As can be readily expected, Si species to form SiC originate from the Si substrate in such a way that holes are formed\cite{Li-JES-142-634,Mogab-JAP-45-1075, Wang-TSF-515-6824, Scholz-DRM-6-1365, Scholz-APA-64-115, Kosugi-APL-74-3939, Severino-JAP-102-2} and only partially observed on the edge of the cross-section of a Si substrate as shown in fig.~\ref{schematic}(c).  That is because the holes are covered by SiC nanowires or films so they are usually unseen.  Secondly, our model proposes that different growth regimes are established due to different distributions of SiC nuclei on Si substrates caused by adjusting the H$_2$/CH$_4$ ratio.  In other words, it is implicitly assumed that SiC nanowires or films are directly grown on a Si substrate depending on the H$_2$/CH$_4$ ratio.  This assumption is supported by the absence of SiC film between a SiC nanowire and a Si substrate, which suggests that SiC nanowires were grown directly from Si substrate, not on SiC film as shown in fig.~\ref{schematic}(d) and (e).

\section{Conclusion}
In conclusion, our study has successfully established the optimal H$_2$ to CH$_4$ gas ratio and growth temperature for synthesizing 3C-SiC nanowires through the carbonization of a Si substrate. These parameters are critical for promoting uniform growth and superior crystalline quality of nanowires while minimizing the formation of amorphous carbon and bulk SiC. By extensively examining the experimental data sets based on kinetics, we have proposed a mechanistic growth model that elucidates the preferential growth of SiC nanowires under high H$_2$/CH$_4$ ratio conditions. These findings not only enhance our understanding of SiC nanowire synthesis but also pave the way for their expanded application in electronics, photonics, and high-temperature environments.

\section{acknowledgement}
This research was supported by Basic Science Research Program through the National Research Foundation of Korea (NRF) funded by the Ministry of Science and ICT (NRF-2021R1A5A1032996, RS-2023-00240724) and through Korea
Basic Science Institute (National research Facilities and Equipment
Center) grant (2021R1A6C101A437) funded by the Ministry of Education.


%

\end{document}